\documentclass[9pt,twocolumn,twoside]{osajnl}
\pdfoutput=1
\journal{optica}
\usepackage{upgreek}
\usepackage[english]{babel}
\usepackage[utf8]{inputenc}
\usepackage{float}
\usepackage{xcolor}
\usepackage{subcaption}
\usepackage{hhline}
\usepackage{amsmath} 
\usepackage{relsize}
\usepackage{array}
\newcolumntype{?}{!{\vrule width 1pt}}
\usepackage[export]{adjustbox}
\setboolean{shortarticle}{false}

\newcommand{\neswarrow}{\mathrel{\text{$\nearrow$\llap{$\swarrow$}}}}
\newcommand{\nwsearrow}{\mathrel{\text{$\nwarrow$\llap{$\searrow$}}}}

\title{Exploiting non-orthogonal eigenmodes in a non-Hermitian optical system to realize sub-100 MHz magneto-optical filters.}

\author[1,*]{Fraser D. Logue}
\author[1]{Jack D. Briscoe}
\author[1]{Danielle Pizzey}
\author[1]{Steven A. Wrathmall}
\author[1]{Ifan G. Hughes}

\affil[1]{Department of Physics, Durham University, Durham, DH1 3LE, United Kingdom}

\affil[*]{Corresponding author: fraser.d.logue2@durham.ac.uk}

\begin{abstract}
Non-Hermitian physics is responsible for many of the counter-intuitive effects observed in optics research opening up new possibilities in sensing, polarization control and measurement. A hallmark of non-Hermitian matrices is the possibility of non-orthogonal eigenvectors resulting in coupling between modes. The advantages of propagation mode coupling have been little explored in magneto-optical filters and other devices based on birefringence. Magneto-optical filters select for ultra-narrow transmission regions by passing light through an atomic medium in the presence of a magnetic field. Passive filter designs have traditionally been limited by Doppler broadening of thermal vapors. Even for  filter designs incorporating a pump laser, transmissions are typically less than 15\% for sub-Doppler features. Here we exploit our understanding of non-Hermitian physics to induce non-orthogonal propagation modes in a vapor and realize better  magneto-optical filters. We construct two new filter designs with ENBWs and maximum transmissions of 181~MHz, 42\% and 140~MHz, 17\% which are the highest figure of merit and first sub-100~MHz FWHM passive filters recorded respectively. This work opens up a range of new filter applications including metrological devices for use outside a lab setting and commends filtering as a new candidate for deeper exploration of non-Hermitian physics such as exceptional points of degeneracy.
\end{abstract}
\setboolean{displaycopyright}{true}

\begin{document}

\maketitle
\section{Introduction}
Non-Hermitian physics \cite{siegman1989excess,berry2004physics,bender2007making,berry2011optical,el2019dawn,el2018non} is opening up new possibilities in photonics from the creation of omnipolarizers \cite{hassan2017chiral,2017PhRvL.118i3002H,2021Optic...8..563K} to sensing applications \cite{rechtsman2017optical,2017Natur.548..192C,2020NatCo..11.2454W} and tailoring laser output \cite{hodaei2014parity,miao2016orbital,gao2017parity}. In general, non-Hermitian matrices do not guarantee orthogonal eigenvectors \cite{brody2013biorthogonal,nowak2018probing,xue2020non} and have real eigenvalues (PT-symmetric) \cite{el2007theory,zyablovsky2014pt,feng2017non}, imaginary eigenvalues (anti-PT-symmetric) \cite{peng2016anti,jiang2019anti,zhang2020synthetic} or complex eigenvalues throughout \cite{wang2020exceptional,luan2022shortcuts}. Ultra narrowband magneto-optical filters \cite{zielinska}, which rely on complex refractive indices, are perfect candidates for non-Hermitian physics but have not yet been studied in this domain. Magneto-optical filters already see many applications in solar weather studies \cite{solar1,giovannelli2020tor,solar4}, laser frequency stabilization \cite{chang2022faradaylaser,tang2022polarization,LaserStab,luo2021,shi2020dual}, LIDAR, \cite{vargas2022mesosphere,xia2020sodium,pan2021metastable}, quantum hybrid systems \cite{QHWidmann}, underwater optical communications \cite{wateroptical} and ghost imaging \cite{yin2022ghost}. Optimizing filters is a balance between maximizing transmission and minimizing bandwidth. For example, tens of \rm{MHz} equivalent noise bandwidth at < $20\%$ transmission has been realized in active filters incorporating a pump beam \cite{Zhuang, Gang,guancold}. On the other hand, several $\rm{GHz}$ equivalent noise bandwidth has been realized at $> 80\%$ transmission in passive designs \cite{speziali2021first,Yin}. A fundamental limit on linewidths has not yet been found and could have future implications for metrology \cite{yan2022freqcomb,auzinsh2022wide,Voigt5}.
\\
\indent Of equal importance is the platform, particularly when filters are designed for use outside a laboratory setting. Most magneto-optical filters rely on thermal vapor cells due to their compact and resilient setups \cite{tut_review}. Thermal vapor filters can be tuned by varying the magnetic field, temperature and input polarization \cite{higgins,Ilja} and can switch between different D-line operation \cite{yan2022dual}. It has been shown that by cascading cells, line-center or wing features can be selected \cite{logue}. However the two setups require the rearrangement of optical elements and the interchange of $\sim$kG permanent magnets with $\sim$100~G solenoids. Note that several metrological applications, such as those that use space, atmospheric or local environmental conditions to detect fluctuations, require apparatus that is robust to the elements \cite{Aerospace1,Aerospace2, wu2019precise}. Therefore there is interest in finding a reconfigurable setup that allows easy switching between wing and line-center operation.
\\
\indent The core principle of a magneto-optical filter is the magneto-optical rotation of light \cite{zhu2021giant,wu2021moems,carothers2022high}. Faraday \cite{liu2022demo,Faraday3, Faraday4} and Voigt \cite{Voigt1F,Voigt3F,Voigt2F} filters, where the magnetic field is parallel or perpendicular to the $k$-vector of the light, rely on circular or linear birefringence respectively. These are special cases where the eigenvectors are orthogonal even though the refractive indices are complex. By setting the magnetic field at an oblique angle to the $k$-vector, the system becomes elliptically birefringent and dichroic \cite{palik1970infrared,papas2014theory}. While oblique filters have been constructed \cite{Voigt2F, rotondaro,higgins,kevfilter}, a treatment on how elliptical birefringence can improve filter performance has not yet been presented. In fact, we show that this elliptical birefringence relies on the eigenmodes of the medium being non-orthogonal. This results in two improvements on the Faraday and Voigt cases: firstly, there is better rejection of light outside the birefringent region and secondly, the birefringent region is much narrower leading to narrower filter peaks and better figures of merit.
\\
\indent In this paper, we theoretically predict the benefits of oblique filters and experimentally realize two competitive filter designs. In Section \ref{Elliptical}, we show how oblique filters can reject light outside of the birefringent region at $>90\%$ efficiency using Stokes parameters to characterise their performance. In Section \ref{Eigenmodes}, we demonstrate that this effect is explained by the oblique geometry having non-orthogonal eigenmodes and introduce the term \textit{invariant polarizations} as the natural additional basis to explain oblique magneto-optical rotation. We calculate the birefringent region showing it to be narrower than the Faraday and Voigt cases. In Section \ref{Experiment}, we realize two thermal vapor filters with lower equivalent noise bandwidth than previously published passive designs. Using naturally abundant Rb vapor, we experimentally realize the first Oblique-Voigt filter, a cascade of an oblique and a Voigt cell, and the first oblique double pass filter, a single oblique cell with light passed through twice. We find excellent agreement with data giving equivalent noise bandwidths and transmissions of 181~MHz, 42~\% and 140~MHz, 17~\% respectively. The first filter is the highest Figure of Merit (FOM) \cite{Ilja} passive filter to date. The second filter is the first sub-100~MHz FHWM passive design and is reconfigurable allowing the selection of line-center or wing features by solenoid current variation.
\vspace{-0.1cm}
\section{Elliptical Birefringence}
\label{Elliptical}
\begin{figure}[!tb]
\begin{center}
{\includegraphics[width=\linewidth]{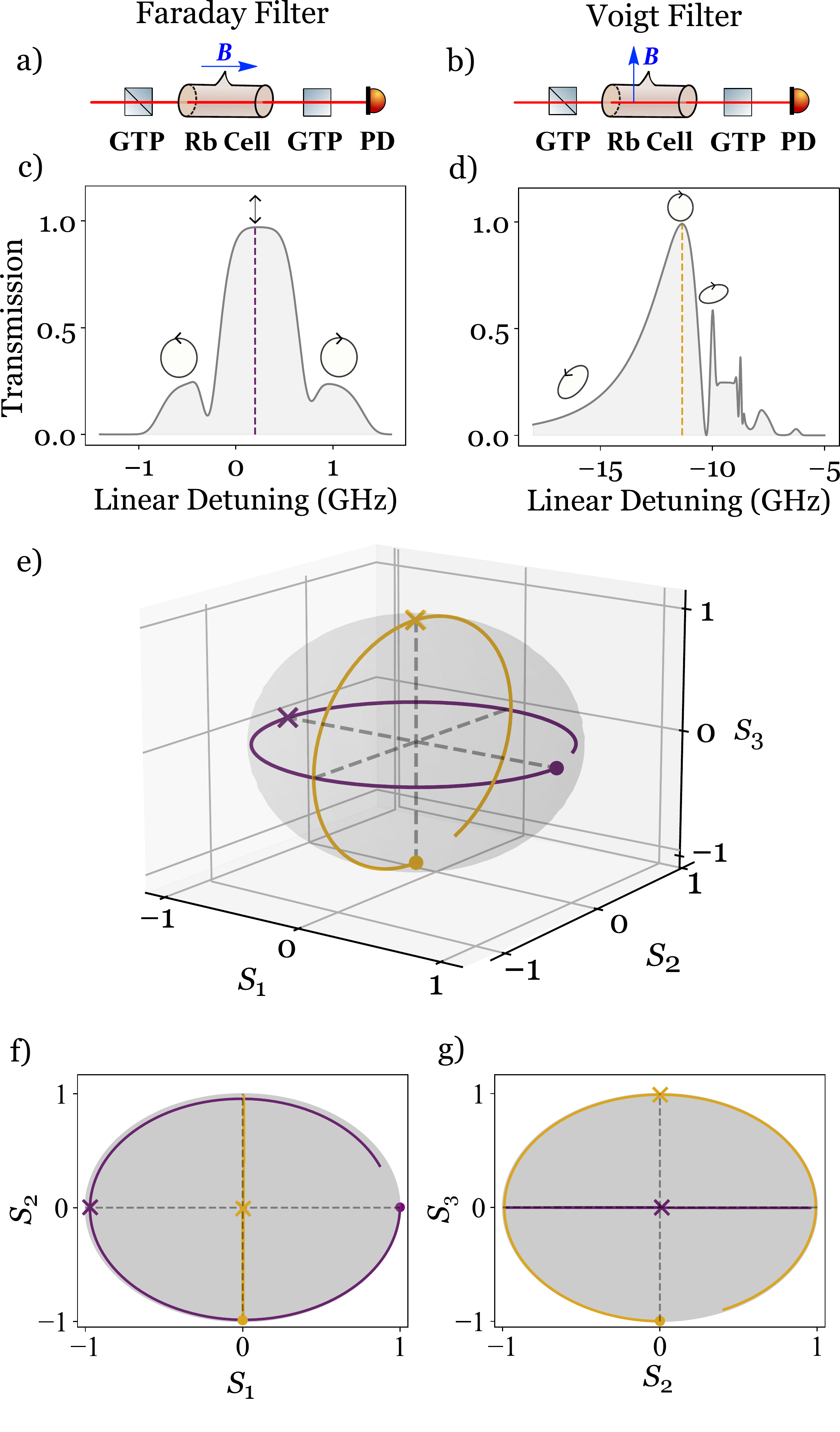}}
\small
\hspace*{0.5cm} 
\begin{tabular*}{\linewidth}{|c|c|c|c|c|c|}
\cline{1-6}
\textbf{Geometry} & $\boldsymbol{\theta\hspace{0.1cm} (^\circ)}$ & \textbf{D-line} & 
 $\boldsymbol{T\hspace{0.1cm}(^\circ\rm{C})}$ & $\boldsymbol{B\hspace{0.1cm}(\rm{G}) }$  & \textbf{Inp. Pol.}\\ \hhline{|=|=|=|=|=|=|}
Faraday & 0 & D2 & 63 & 160 & Lin. Hor. $\leftrightarrow$\\
Voigt & 90 &D1 & 100 & 3500 & L.H. Circ. $\circlearrowleft$\\
\cline{1-6}
\end{tabular*}
\end{center}
\caption{a) and b) : Schematics of single cell Faraday and Voigt filters in natural abundance Rb. Note the second polarizer is rotated so that the hypotenuse is not in the plane of the page. As such the polarizers are crossed. Parameters shown in the table. c) and d): The filter transmissions after 75~mm plotted against linear detuning with zero detuning being the weighted centre of the resonance frequencies of the two Rubidium isotopes. Colored dotted lines indicate the frequencies plotted on the Poincar\'e sphere. Polarization animations shown. e): A Poincar\'e sphere representation showing the polarization evolution of light moving through the Rb vapor cell. 'x' marks the evolution after 75~mm. f) and g) are projections on to the $S_3$ and $S_1$ planes respectively. Stokes parameters are normalized with respect to intensity of input light. The Faraday maximum transmission frequency evolves on the equator while the Voigt maximum transmission frequency evolves along a meridian.}
\label{fig:poincare}
\end{figure}

\begin{figure*}[!tb]
\includegraphics[width=\linewidth]{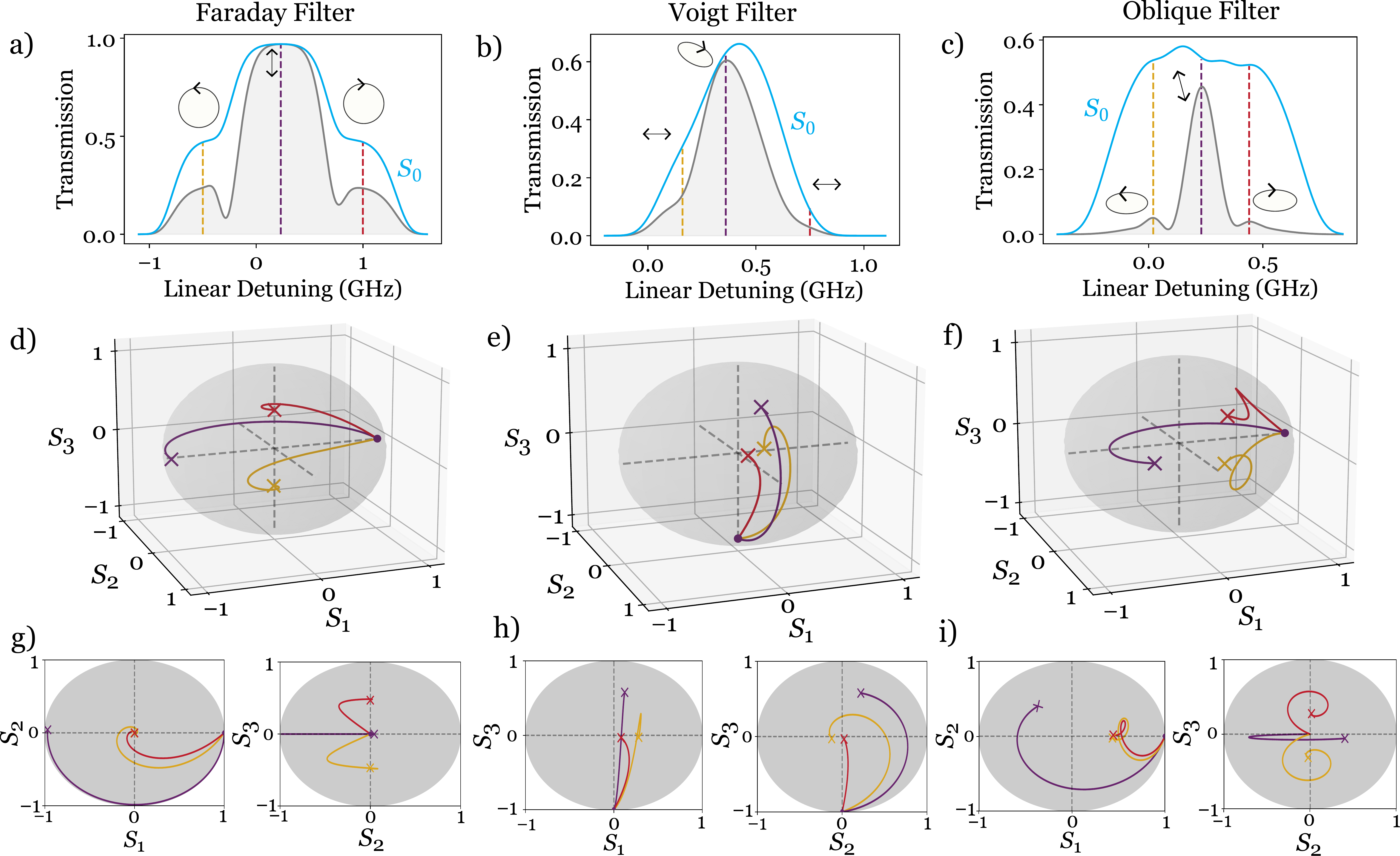}
\hspace*{1cm} 
\begin{tabular*}{\linewidth}{|c|c|c|c|c|c|c|c|}
\cline{1-8}
\textbf{Geometry} & $\boldsymbol{\theta \hspace{0.1cm} (^\circ)}$ & \textbf{D-line} & 
 $\boldsymbol{T\hspace{0.1cm}(^\circ\rm{C})}$ & $\boldsymbol{B\hspace{0.1cm}(\rm{G}) }$ & \textbf{Inp. Pol.} &  \textbf{\% Rejection (Gold Freq.)} & \textbf{\% Rejection (Red Freq.)} \\ \hhline{|=|=|=|=|=|=|=|=|}
Faraday & 0 & D2 & 63 & 160  & Lin. Hor. $\leftrightarrow$ & 49 & 50\\
Voigt & 90 & D1 & 107 & 514  & L.H. Circ. $\circlearrowleft$  & 54 & 68\\
Oblique & 86 & D2 & 99 & 226  & Lin Hor. $\leftrightarrow$   & 96 & 92\\
\cline{1-8}
\end{tabular*}
    \caption{a), b), c): Theory plots of Faraday, Voigt and oblique filter transmissions in 75~mm naturally abundant Rb vapor cells (gray) plotted alongside $S_0$ (blue). Parameters given in table. We choose to plot the D2 line for the Faraday and oblique filters for comparison. We plot the D1 line instead for the Voigt filter as similar line-center D2 filters are not competitive in terms of FOM with the Faraday and oblique examples.  The filter maximum transmission frequency (purple dashed line) and two frequencies outside the birefringent region (yellow- and red-dashed line) are marked.  Note the red Voigt frequency remains slightly left handed after propagation leading to the > 50\% rejection.  Animations depict the output polarization before the second polarizer. d), e), f): Poincar\'e sphere representations of the three frequencies. Stokes parameters are normalized with respect to input intensity. g), h), i): Projections of the Poincar\'e sphere on to different axes. We note that output elliptical horizontal states either side of the oblique filter peak result in better filter rejection than the circular and linear horizontal states in the Faraday and Voigt filters respectively leading to improved filter performance.}
    \label{fig:poincare_oblique_faraday_voigt}
\end{figure*}

The orientation of the magnetic field defines the relationship between input polarization and the induced $\pi$ ($m_J = m'_J$) and $\sigma^{+/-}$ ($m_J = m'_J \pm 1$) electric dipole allowed transitions \cite{fran,rotondaro,foot2004atomic}. In the Faraday geometry, a magnetic field is exerted parallel to the $k$-vector of the input light which we take to be parallel with the $z$-axis. This results in left/right handed circular light inducing $\sigma^{+/-}$ transitions \cite{adams2018}. In the Voigt geometry, the magnetic field is perpendicular to the $k$-vector and we have a choice of field lines running parallel to the $y$- or $x$-axis. Without loss of generality, in this paper, we always place magnets in the $x-z$ plane (i.e. on the surface of an optical bench). In this case, horizontally polarized light induces $\pi$ transitions and vertically polarized light induces a linear combination of $\sigma^+$ and $\sigma^-$ transitions \cite{Voigt4}.
\\
\indent Differences in the transition probabilities lead to polarization dependent absorption (dichroism) and continuous polarization state transformation (birefringence) as the light propagates through the vapor \cite{sargsyan,shang2022measurement}. The Faraday geometry exhibits a circular birefringence as differential refraction of circular handed light causes magneto-optical rotation of the plane of linear light. In contrast, the Voigt geometry exhibits linear birefringence where differential refraction of horizontally and vertically polarized light leads to the handedness cycling through right, linear and left hands.
We can describe polarization evolution using Stokes parameters \cite{schaefer2007measuring,goldberg2021quantum} defined as,

\begin{equation}
S_0 = \frac{I_{\rm{out}}}{I_{\rm{in}}},
\end{equation}
\vspace{0.01cm}
\begin{equation}
S_1 = \frac{(I_\leftrightarrow-I_\updownarrow)}{N},\hspace{0.2cm}
S_2 = \frac{(I_{\neswarrow} - I_{\nwsearrow})}{N},\hspace{0.2cm}
S_3 = \frac{(I_{\mathlarger{\circlearrowright}} - I_\mathlarger{\circlearrowleft})}{N}.
\end{equation}

\noindent $S_0$ represents the total transmission and is proportional to $I_{\rm{out}}$, the output intensity. $S_1$, $S_2$ and $S_3$ together uniquely describe the polarization state. The arrow indices depict the polarization state of the output light ($\circlearrowright/\circlearrowleft$ denotes right/left hand circular light) and $N$ is a normalization factor. Throughout when we write $S_1$, $S_2$ and $S_3$, we are using input normalized stokes vectors $(N=I_{\rm{in}})$. When we write tildes, $\tilde{S}_1,\tilde{S}_2,\tilde{S}_3$, we are denoting output normalized stokes vectors $(N=I_{\rm{out}})$.
\\
Plotting Stokes parameters on a Poincar\'e sphere provides a useful way of visualizing polarization \cite{Ilja,dantan2006spin,weller2012measuring,su2021measuring}. In this paper, we explain filter performance by plotting the Stokes parameters as a function of the distance light has propagated through the vapor. A magneto-optical filter only outputs light that has rotated to be orthogonal to the input polarization. This is represented on the Poincar\'e sphere when curves  travel to the antipode of the input (assuming no loss).  Fig. \ref{fig:poincare} plots evolutions of maximum transmission frequencies for a Faraday and  Voigt filter demonstrating circular birefringence, where the polarization evolves on the equator, and linear birefringence, where the polarization evolves along a meridian.

If a magnetic field is applied at an angle $\theta$ to the $k$-vector, the projection operators for the three transitions are:

\begin{equation}
\begin{split}
    &\sigma^+: \frac{\widehat{x}\cos\theta+ \mathrm{i}\widehat{y}}{{\sqrt{2}}}. \\
    &\sigma^-: \frac{\widehat{x}\cos\theta- \mathrm{i}\widehat{y}}{{\sqrt{2}}}.\\
    &\pi: \hspace{0.2cm} \widehat{x}\sin\theta.
\end{split}
\label{projection operators}
\end{equation}
The $\pi$ transition operator is always horizontally linear. If $\theta$ is an oblique angle, neither $0$ or $90^\circ$ to the $k$-vector, the $\sigma$ transition operators are elliptical. Therefore, the oblique geometry exhibits elliptical birefringence where we have differential refraction between elliptical polarization states \cite{palik1970infrared,papas2014theory}. A derivation of the projection operators can be found in Appendix A of the Supplementary Material.
\\
In atom-light interactions, there is a birefringent region, where magneto-optical rotation occurs. In a Faraday/Voigt birefringent region, circular/linear birefringence results in linear/circular light as an output. Outside of the birefringent region, circular/linear dichroism of light outputs circular/linear light. Given that circular and linear light are not orthogonal polarization states, a filter cannot completely reject the output light outside the region but filters only $\sim50\%$. However, by choosing a suitable value of $\theta$, it is possible to construct an oblique filter where light outside the birefringent region is suppressed beyond >~90\%. Fig. \ref{fig:poincare_oblique_faraday_voigt} compares the filter transmissions of Faraday, Voigt and oblique filters with $S_0$. 
\\
Note that the Faraday filter has characteristic pedestals either side of the central peak where circular light is rejected at 50\%. In the Voigt filter, we also see $\sim$~50\% rejection for the selected negatively detuned frequency (gold). The positively detuned frequency (red) is rejected at >~50\% but the S0 value is already less than 10\%. However, the oblique filter demonstrates unparalled rejection of light outside the birefringent region significantly reducing the equivalent noise bandwidth, (ENBW),

\begin{equation}
 \mathrm{ENBW}=\int\frac{\mathcal{T}(\nu)\mathrm{d}\nu}{\mathcal{T}(\nu_{\rm{s}})},
\end{equation}

\noindent where $\mathcal{T}(\nu)$ is the transmission for a given frequency, $\nu$, or the maximum transmission frequency, $\nu_s$. In the literature, a Figure of Merit (FOM) \cite{Ilja},

\begin{equation}
 \mathrm{FOM}=\frac{\mathcal{T}({\nu}_{\rm{s}})}{\mathrm{ENBW}},
\end{equation}

\noindent is often used to compare filter quality with a lower ENBW increasing FOM.

\begin{figure*}[!tb]
\centering
{\includegraphics[width=\linewidth]{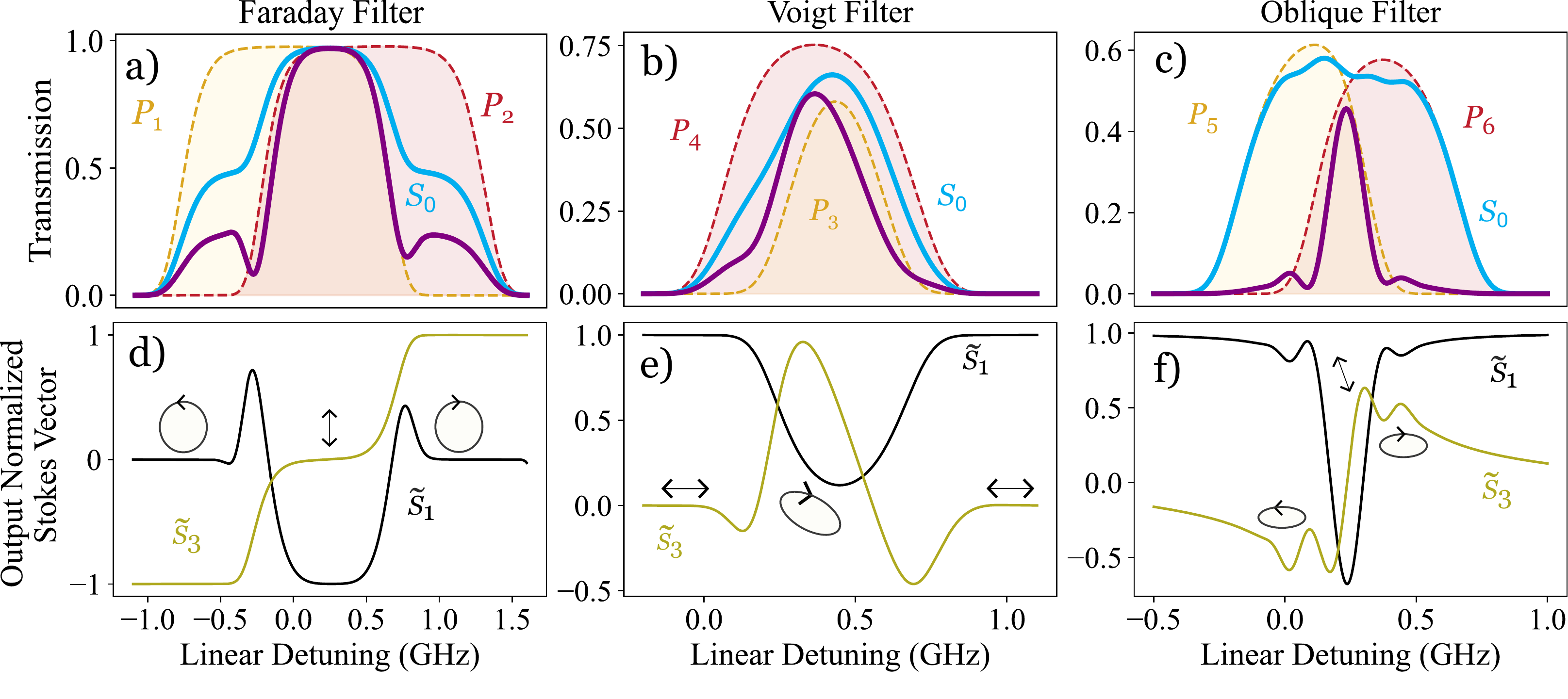}}
\hspace*{0.7cm} 
\begin{tabular*}{\linewidth}{|c|c|c|c|c|c|c|c|}
\cline{1-8}
\textbf{Geometry} & $\boldsymbol{\theta (^\circ)}$ & \textbf{D-line} & 
 $\boldsymbol{T\hspace{0.1cm}(^\circ\rm{C})}$ & $\boldsymbol{B\hspace{0.1cm}(\rm{G}) }$ & \textbf{Inp. Pol.} & \textbf{Birefringent Region FWHM (MHz)} & \textbf{Filter FWHM (MHz)} \\ \hhline{|=|=|=|=|=|=|=|=|}
Faraday & 0 & D2 & 63 & 160  & Lin. Hor.$ \leftrightarrow$  & 850 & 790\\
Voigt &  90 & D1 & 107 & 514  & L.H. Circ. $\circlearrowleft$  & 320 & 310\\
Oblique & 86 & D2 & 99 & 226  & Lin Hor.$ \leftrightarrow$  & 200 & 150\\
\cline{1-8}
\end{tabular*}
\caption{a), b), c): Theory plots of Faraday, Voigt and oblique filter transmissions in 75~mm naturally abundant Rb vapor cells (purple) plotted alongside $S_0$ (blue). Parameters given in table. The red and yellow regions are the $S_0$ output if the two invariant polarizations are input (see Table \ref{Table}). In the Faraday and Voigt cases, these correspond to the transmission regions left after absorption due to one or more transitions. In the oblique case, the invariant polarizations are frequency dependent. The birefringent region is shown in orange as the overlap between the red and yellow regions. In the Voigt case, one transmission region ($P_3$) completely overlaps with the other ($P_4$). d), e), f) depict the $\rm{\tilde{S}_1}$ and $\rm{\tilde{S}_3}$ Stokes parameters for the three filters with polarization animations for the same frequencies as in Fig. \ref{fig:poincare_oblique_faraday_voigt}. Note the Stokes vectors are normalized with respect to the output intensity and therefore give information about polarization state but not intensity loss. The narrowness of this oblique birefringent region results in a FWHM over two times and five times narrower than the Voigt and Faraday filter respectively.}
\label{birefringent regions}
\end{figure*}

\section{Birefringent Regions with Non-Orthogonal Eigenmodes}
\label{Eigenmodes}
We can learn more about the magneto-optical rotation of a vapor by calculating the birefringent region explicitly. The birefringent region is the $S_0$ output where differential refraction occurs.  Better filter peaks are formed within birefringent regions since they are highly selective resulting in narrower lineshapes and can rotate light with less loss. A filter may have additional transmission due to absorption both inside and outside of this region. Dichroic absorption alone evolves the polarization to one particular polarization state after which point moving through more vapor reduces the intensity of the light without altering the polarization state. This is a high loss process and as discussed in Section \ref{Elliptical}, light in this region is rejected by the filter at > 50\%.

To compute the birefringent region we need to find a basis composed of polarizations which do not transform as they move through the vapor. We call these basis vectors \textit{invariant polarizations}. In the Faraday and Voigt geometries, the invariant polarizations are the propagation eigenmodes derived from solving the wave equation \cite{rotondaro}. We then calculate two $S_0$ regions, shown in red and yellow in Fig. \ref{Eigenmodes}, with the two eigenmodes as input polarizations. The overlap of the two regions is the birefringent region, shown in orange. In this detuning range, any input polarization with non-zero components of both eigenmodes undergoes continuous polarization state transformation.

However in the oblique geometry, the eigenmode solutions to the wave equation are non-orthogonal and frequency dependent. As such the eigenmodes are not invariant polarizations. Instead, we find the polarizations orthogonal to each eigenmode which do not undergo differential refraction. For more details, see Appendix B of the Supplementary Material. The full list of invariant polarizations is shown in Table \ref{Table}. In particular, note that the oblique invariant polarizations generally induce all three transitions provided the three frequency dependent transition probabilities are non-zero.

\setcounter{table}{2}
\begin{table}[H]
\begin{tabular}{|c|c|c|}
\cline{1-3}
\textbf{Transmission Region} & \textbf{Invariant Pol.} & \textbf{Transitions Ind.}\\ \hhline{|=|=|=|}
$P_1$ (Faraday) & L.H.C $\circlearrowleft$ & $\sigma^+$ \\
$P_2$ (Faraday) & R.H.C. $\circlearrowright$ & $\sigma^-$ \\
$P_3$ (Voigt) & Lin. Hor. $\leftrightarrow$ & $\pi$ \\
$P_4$ (Voigt) & Ver. Hor. $\updownarrow$ & $\sigma^-/\sigma^+$ \\
$P_5$ (Oblique) & Freq. Dep. & $\sigma^-/\sigma^+/\pi$ \\
$P_6$ (Oblique) & Freq. Dep &  $\sigma^-/\sigma^+/\pi$\\
\cline{1-3}
\end{tabular}
\caption{A table of the 6 $S_0$ regions shown in Fig. \ref{Eigenmodes}. Each transmission region is created by inputting an invariant polarization to the cell which does transform polarization state upon propagation i.e. is not magneto-optically rotated. Note that in the Faraday and Voigt cases, the invariant polarizations are the eigenmodes of the system and can be read off from their respective angular dipole matrix elements (See Appendix A of the Supplementary Material). In the oblique case, the invariant polarizations are frequency dependent. The invariant polarizations induce certain transitions and these regions are the transmission left after transition induced absorption}
\label{Table}
\end{table}
\vspace{0.2cm}
In Fig. \ref{birefringent regions}, we observe that the width of the birefringent region bounds the width of the filter peak. In the Faraday and oblique cases this is easier to see. In the Voigt case, one transmission region ($P_3$) lies entirely within the other ($P_4$). The larger transmission of $P_4$ results in the center of the peak being displaced and some absorption outside the birefringent region contributing to the peak's profile. The plotted oblique filter has the narrowest birefringent region resulting in the narrowest FWHM peak, two times and five times narrower than the Voigt and Faraday filters respectively.

Summarizing Sections \ref{Elliptical} and \ref{birefringent regions}, we can give a complete description of why the oblique filter can be tuned to give the best filter. Given that the invariant polarizations induce all three transitions, the $S_0$ regions and hence their overlap (the birefringent region) is narrower leading to narrower peaks. Outside of this region, dichroic absorption results in transformation to polarization states which can be filtered at >~90\%. As such, the filter profile is improved as most light is transmitted within the birefringent region. The ENBW is also lower for this reason and additionally the FWHM of the peak is narrower. The detriment to maximum transmission does not outweigh the reduction in ENBW resulting in higher FOMs.

\section{Experiment and Results}
\label{Experiment}
\begin{figure}[!tb]
\centering
{\includegraphics[width=\linewidth]{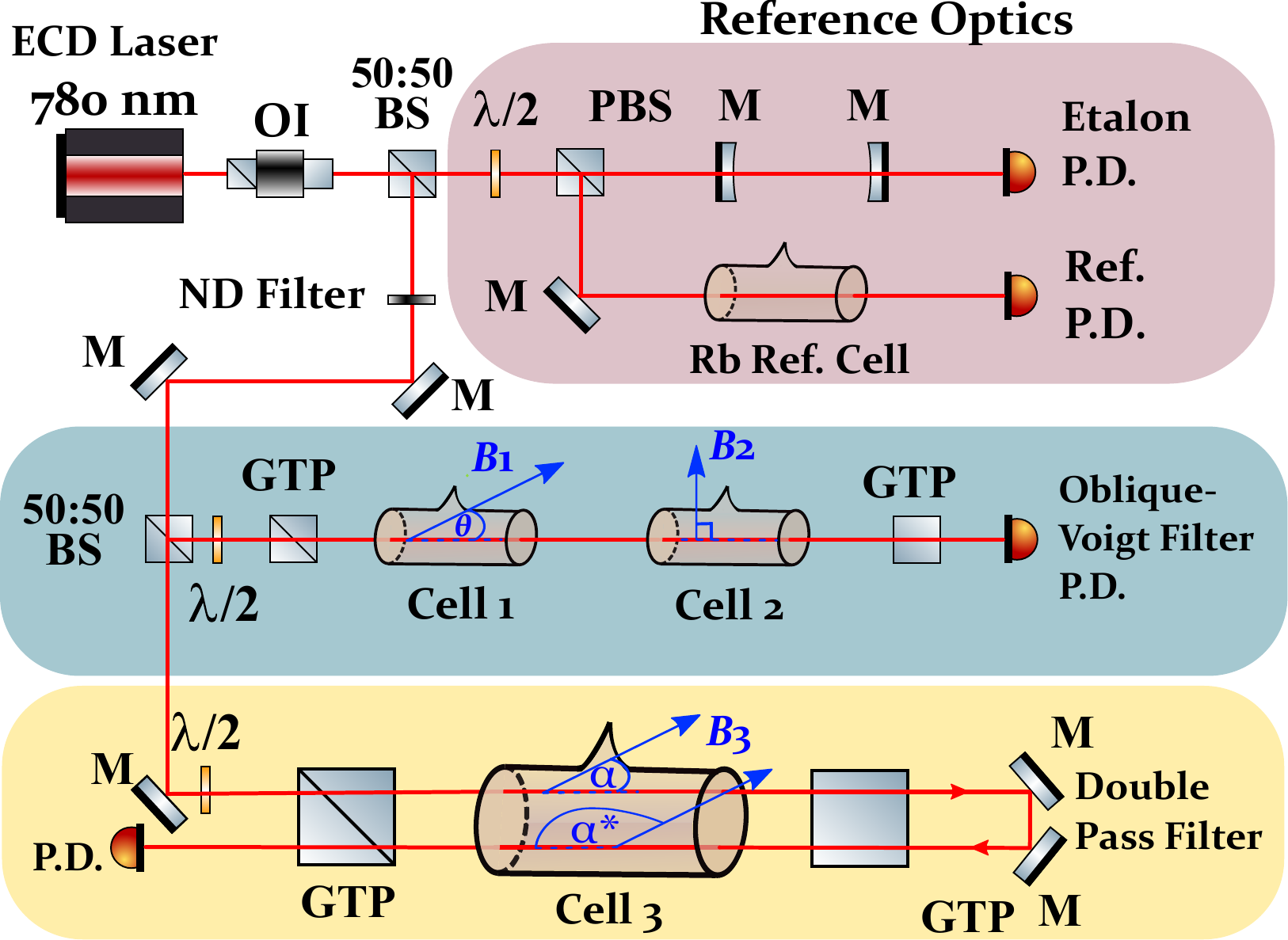}}
\caption{An illustration of the experiment. An external cavity diode (ECD) laser tuned to the Rb-D2 line emits light towards reference optics and the two experiments. The reference optics include a Fabry-P\'erot etalon to linearize the scan and a rubidium reference (Rb Ref.) cell for absolute frequency calibration. A neutral density (ND) filter lowers the intensity of the light entering the experiment. $\lambda/2$ waveplates rotate the light into the correct input polarization. The Oblique-Voigt filter consists of two rubidium vapor cells between two Glan-Taylor polarizers (GTP) which are crossed. Cell 1 has a magnetic field exerted at an angle $\theta$ to the $k$-vector while the magnetic field is perpendicular to the $k$-vector for Cell 2. The double pass filter consists of one cell between crossed Glan-Taylor polarizers for each pass. On the first pass the magnetic field makes an angle $\alpha$ to the $k$-vector and on the second pass makes an angle $\alpha^*=(180 - \alpha)^\circ$. Cells 1 and 3 have magnetic fields generated via a combination of cuboidal plate magnets and solenoids which exert the appropriate transverse and axial fields respectively. Cell 1 and 3 are heated using adhesive flexible heating elements stuck along the length of the cells. Cell 2 is placed in a copper heater and has a transverse magnetic field generated by NdFeB top hat magnets \cite{brown2002developments,weller2012absolute}. M- Mirror, OI - Optical Isolator, (P)BS - (Polarizing) Beamsplitter, PD - Photodetector.  }
\label{fig:setup}
\end{figure}

\begin{figure}[tb!]
\centering
\vspace{1cm}
{\includegraphics[width=\linewidth]{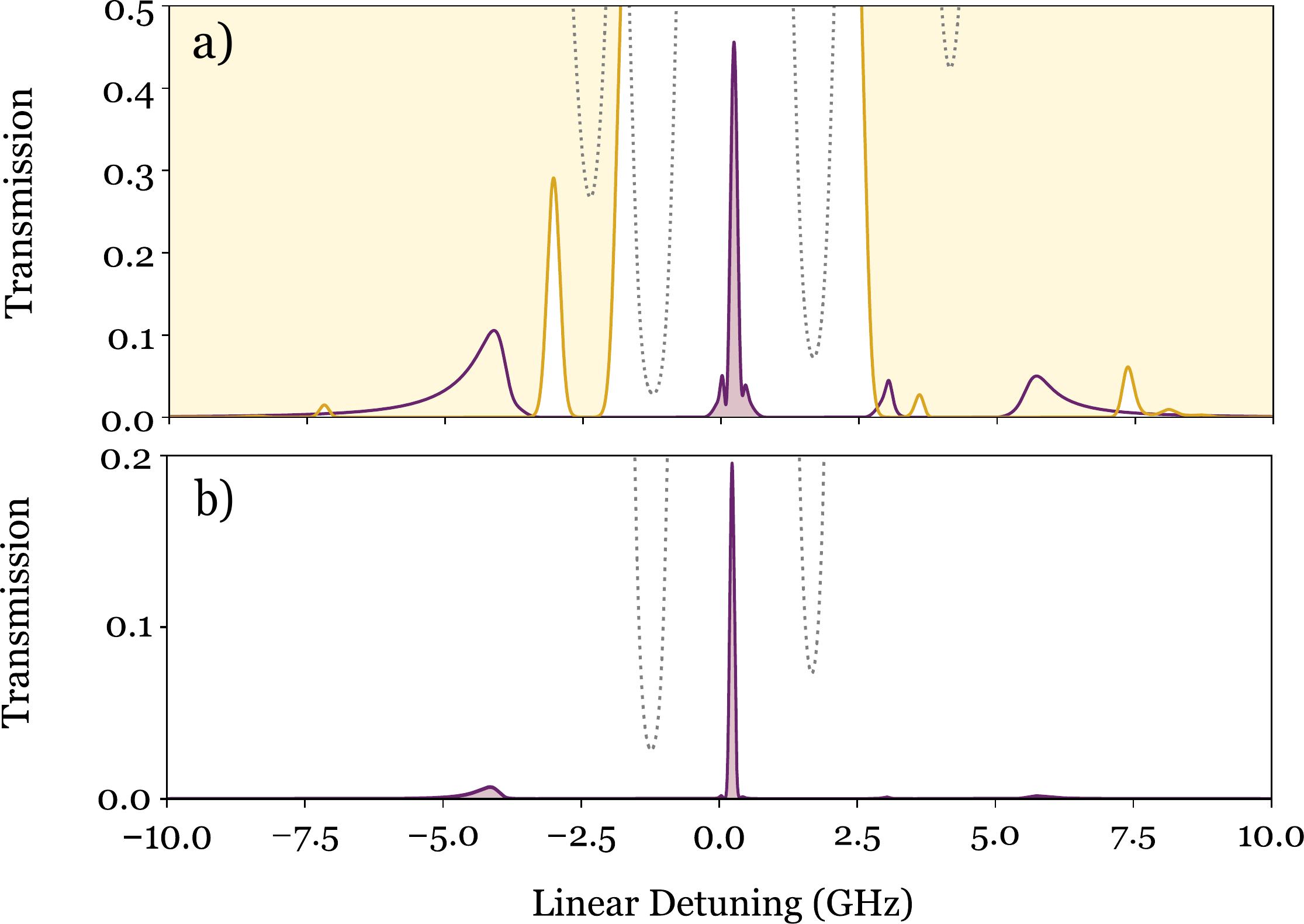}}
\caption{a) The output of the Oblique-Voigt filter with the second cell removed (purple line) and with the second cell (purple shading). The transmission of the second cell (yellow line) with its absorption region (yellow shading) is shown. b) The output of the oblique double pass filter (purple). A zero field natural abundance Rb-D2 spectrum at $35^\circ\rm{C}$ is plotted on both panels (dotted gray).}
\label{fig:cascades}
\end{figure}

\begin{figure*}[tb]
\centering
{\includegraphics[width=0.5\linewidth]{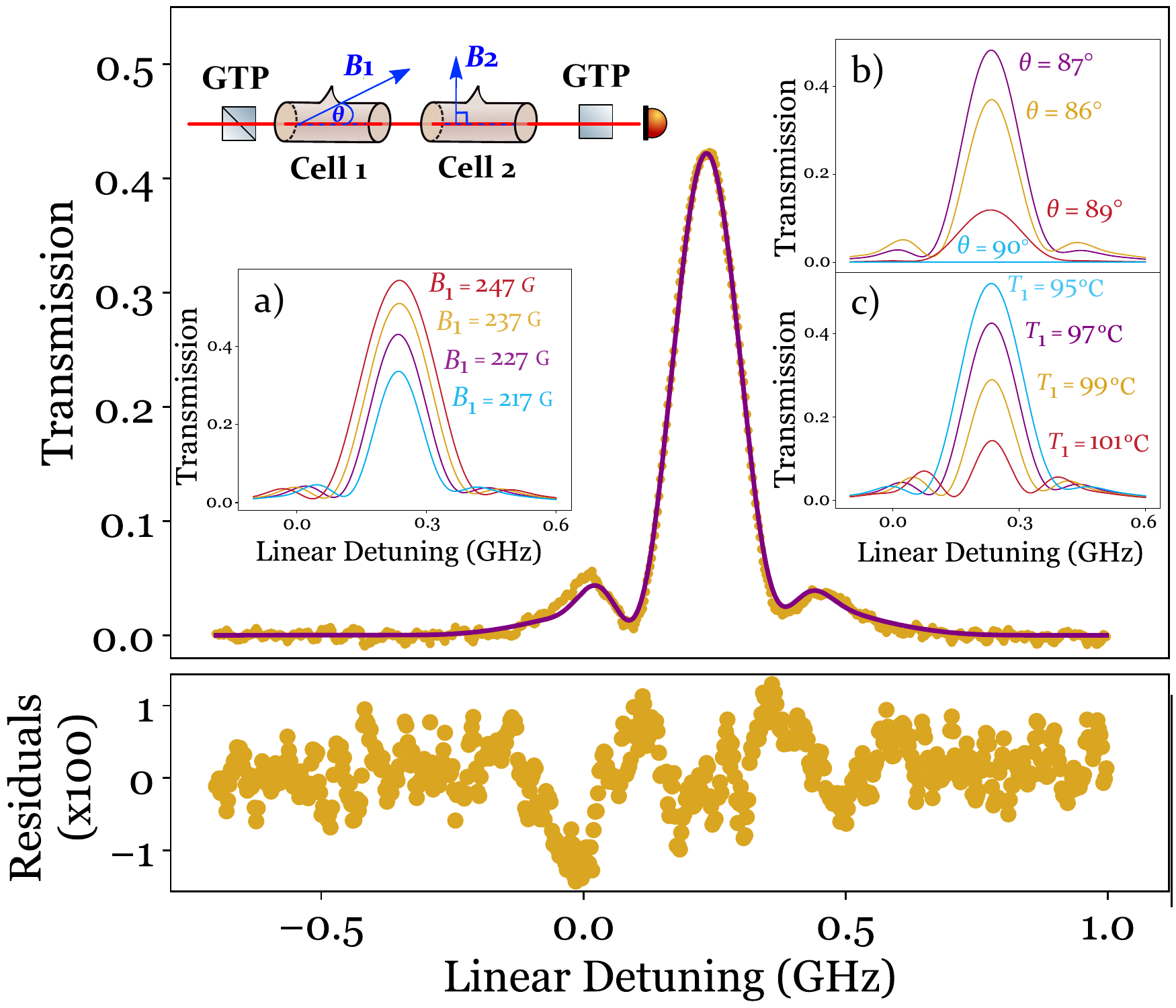}} \hspace{0.03cm} \hfill
{\includegraphics[width=0.49\linewidth]{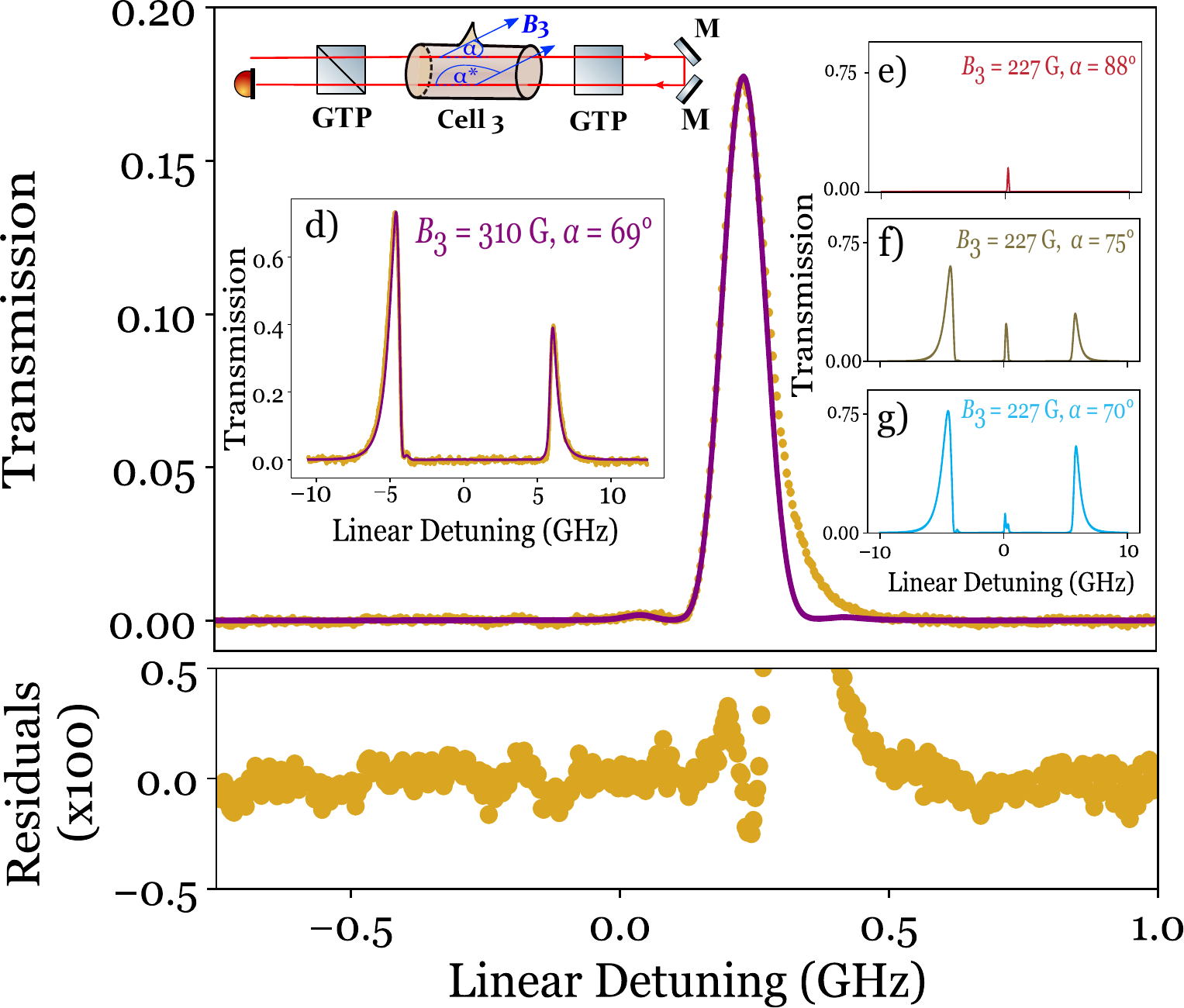}}\hspace{0.01cm}
\begin{center}
\begin{tabular*}{1.5\linewidth}{|c|c|c|c|c|c|c|c|c|}
\cline{1-9}
\textbf{Filter Type} & $\boldsymbol{T_1\hspace{0.1cm}(^\circ\rm{C})}$ & $\boldsymbol{B_1\hspace{0.1cm}(\rm{G}) }$ &
$\boldsymbol{\theta/\alpha \hspace{0.1cm}(^\circ)}$ & $\boldsymbol{T_2\hspace{0.1cm}(^\circ\rm{C})}$ &
 $\boldsymbol{B_2\hspace{0.1cm}(\rm{G}})$ & \textbf{ENBW} $\boldsymbol{(\rm{GHz})}$ & \textbf{FWHM} $\boldsymbol{(\rm{MHz})}$ & \textbf{FOM} $\boldsymbol{(\rm{GHz^{-1}})}$  \\ \hhline{|=|=|=|=|=|=|=|=|=|}
Oblique-Voigt & $96.8~\pm~0.1$ & $226.8 \pm 0.7$ & $86.3 \pm 0.3$ & $123.6~\pm~0.2$ & $3897 \pm 3$ & $181 ~\pm ~1$ & $145 ~\pm ~1$ & $2.38 ~\pm ~0.01$ \\ \cline{1-9}
Double Pass & $99.4~\pm~0.4$ & $226.3\pm 1.2$ & $87.4\pm 0.3$ & $/$ & $/$ & $140~\pm 1$ & $93~\pm 1$ & $1.20~\pm ~0.01$ \\
\cline{1-9}
\end{tabular*}
\end{center}
\caption{(Left) Main plot shows data (gold) and theory (purple) plotted for a natural abundance Rb-D2 two cell cascade with 75~mm and 5~mm cells in the weak probe regime. The data show excellent agreement with theory with RMS fit error of 0.5\%. The insets show theory spectra where  a)\hspace{0.05cm}$B_1$, b)\hspace{0.1cm}$\theta$ and c)\hspace{0.05cm} $T_1$ are varied. All other parameters are fixed. (Right) Main plot shows data (gold) and theory (purple) for a natural abundance Rb-D2 double pass filter in the weak probe regime. The data show excellent agreement with theory apart from a shoulder with RMS fit error of 0.3\%. d) shows an experimental wing filter with fit achieved using the same double pass setup by increasing the current in the solenoid and decreasing the separation of the cuboidal magnets. This changes $\alpha$ to $69^\circ$ and $B$ to $310$~G, the temperature has also lowered to $92^\circ \rm{C}$ e), f), g) show the theoretical evolution of the filter's features as the angle $\alpha$ is varied. The table shows the mean parameter values obtained from fits of five spectra. The ENBW, FWHM and FOM fit values also account for the systematic errors involved in data acquisition and linearization.}

\label{fig:exp1}
\end{figure*}

In order to test the advantages of the oblique geometry we realized two filter designs; the experimental setup is shown in Fig. \ref{fig:setup}. In the experimental arm, light on the Rb-D2 line with a $1/e^2$ beamwidth of $100~\upmu \rm{m}$ passes through an ND filter to lower the power to 300~nW. We work in the weak probe regime \cite{weak, bala2022doppler,luo2018signal} as it is readily available to model and higher intensity filters are outside the scope of this work.
\\ The aim of the two filter designs is to output the desirable central filter features studied in previous sections while eliminating features away from center. Fig. \ref{fig:cascades} shows how the two designs eliminate unwanted features over a 20~GHz detuning range. The first experiment is a two cell cascade with the first cell in the oblique geometry and the second in the Voigt geometry. Both cells are between crossed Glan-Taylor polarizers. The two cells play different roles: Cell 1's role is to create the narrow feature at center as studied in the previous sections while the role of Cell 2 is to absorb features away from the central region. A Voigt cell is particularly appropriate for this role since the magnetic field and temperature can be tuned to be highly absorptive away from center leaving a small transmission region at center for the narrow peak to be transmitted through. The theory behind the Voigt cell has been studied in \cite{logue}. This is the first Oblique-Voigt cascade recorded in the literature.
\\
The second experiment is a double pass filter where light is passed through an oblique cell between polarizers and filtered before being reflected back into the cell and filtered again.  The rays from the two passes follow a similar path and diverge by less than a 1~mm in the $x-y$ plane over the cell length. As the two beams are weak, neither beam pumps the atoms into higher states.
\\
On the first pass, the light input is horizontally linear polarized and the magnetic field makes an angle $\alpha$ to the propagation direction while on its second pass the light is filtered to be vertically linear polarized and the angle of the magnetic field is $\alpha^*=(180-\alpha)^\circ$. The two passes individually give the same filter output. Left/right handed light propagating at an angle $\alpha$ to the magnetic field is equivalent to right/left handed light propagating at an angle $\alpha^*$, however given we are inputting linear light the interchange of $\alpha^*$ for $\alpha$ results in no change. Additionally, horizontal and vertical input always gives the same filter output, provided the polarizers are rotated appropriately, regardless of the magnetic field angle (see Appendix C of the Supplementary Material).  As such, filtering twice results in a squared intensity effect which suppresses the features away from center much more than the central features present after the first pass. Double pass filters have been realized before \cite{luofilter,turner2002sub,ray2013study,zhu2017theoretical} however never with an oblique cell.
\\
Using an adapted version of the open-source software package \textit{ElecSus} \cite{elecsus1,elecsus2}, we found parameters for natural abundance Rb that would give a large FOM for the Oblique-Voigt filter and narrow FWHM for the double pass filter. We found that both of these objectives can be obtained using very similar parameters for Cell 1 and Cell 3. We chose 75mm cells for Cell 1 and Cell 3 and a 5mm cell for Cell 2. Choosing cells is an interplay between achieving a homogeneous magnetic field which is easiest in a small cell and avoiding collisional broadening effects which occur at higher temperatures. Hence if a large number density \cite{wu1986optical} is needed, longer cells may be better \cite{ZentileBroad}. The combination of a solenoid and permanent magnets allows us to use a longer cell than in previous oblique investigations \cite{kevfilter,higgins} since the magnitude of the field is largely provided by the permanent magnets while its direction is determined by small adjustments in the solenoid current. This allows easier control of the field magnitude and direction with the permanent magnets set square with the cell resulting in less error in homogeneity.
\\
The results and fit parameters are shown in Fig. \ref{fig:exp1}. The data show excellent agreement with theory \cite{Hughes2010} apart from a shoulder on the double pass filter. The Oblique-Voigt filter has the highest FOM recorded to date of $2.38\pm~0.01~ \rm{GHz}^{-1}$. The double pass has the highest FOM of a single cell passive filter recorded to date, $1.20\pm~0.01~ \rm{GHz}^{-1}$,  and has a sub-100~MHz fit FWHM of 93~MHz.
\\
The insets on the left graph show the dependence of the Oblique-Voigt filter performance on $\theta$, $B_1$ and $T_1$. We note the high sensitivity to $\theta$. If $\theta=90^\circ$, there is no filter transmission since in the Voigt geometry linear horizontal light is not rotated. However as $\theta$ decreases, a filter profile quickly emerges showing the radical difference in the birefringence between the Voigt and oblique cases.
\\
The insets on the right graph show how the double pass filter can be reconfigured to select for wing features over the line-center feature. This is achieved by raising the current which increases the axial field reducing the angle $\theta$ in the resultant field. We experimentally realize a wing filter using this procedure with the cuboidal magnets placed slightly closer to raise the magnitude of the magnetic field to $\sim 300~G$. Such a feature would be very useful in a manufactured device where it may not be feasible to rearrange optical elements to achieve the same result.

\section{Conclusion}
In conclusion, we have presented the underlying non-Hermitian physics behind filters in the oblique geometry. We have shown that elliptical birefringence and non-orthogonal propagation eigenmodes can create narrower filters with greater rejection of light outside the birefringent region. We realized two new filter designs: an Oblique-Voigt cascade and an oblique double pass which showed excellent agreement with theory. These are the highest FOM and first sub-100 MHz FWHM passive design filters recorded respectively. These findings alongside the reconfigurability and compactness of the setup are promising for future applications in metrology outside a lab context 
\cite{chang2022integrated,picque2019frequency}. This is an initial work into non-Hermitian filter effects; our theory \textit{ElecSus} predicts frequencies where the non-orthogonal eigenmodes almost completely coallesce suggesting the presence of exceptional points of degeneracy \cite{heiss2012physics,heiss2016circling,ozdemir2019parity,miri2019exceptional}. In future work, we want to consider the additional sensing capabilities a device would have owing to exceptional points' $\sqrt{\epsilon}$ sensitivity for small parameter $\epsilon$ \cite{2017Natur.548..192C,2020NatCo..11.2454W}.
\begin{backmatter}
\bmsection{Funding} ESPRC Grant EP/R002061/1

\bmsection{Acknowledgments} We would like to thank Thomas Robertson-Brown for their construction and characterization of the magnetic designs and Sharaa Alqarni for the use of their reference optics.

\smallskip

\bmsection{Disclosures} The authors declare no conflicts of interest.

\bmsection{Data availability} Data underlying the results presented in this paper are available in Ref. \cite{data}.
\vspace{-0.3cm}
\end{backmatter}
\bibliography{main}

\section*{Appendix A: Deriving the Transition Projection Operators in the Oblique Geometry}

We define $z$ as the propagation direction of the light and note that the light is polarized transversely in the  $x-y$ plane. In the case where $z$ is the quantization axis, the angular part of the dipole matrix element is: 

\begin{equation}
   \mathcal{M}= \sqrt{\frac{4\pi}{3}}\langle l'm_l'|\frac{\widehat{x} + \mathrm{i}\widehat{y}}{\sqrt{2}}Y_{1,1}+\frac{\widehat{x} - \mathrm{i}\widehat{y}}{\sqrt{2}}Y_{1,-1}+\widehat{z}\hspace{0.1cm}Y_{1,0}|lm_l\rangle,
\label{eq}
\end{equation}

\noindent where $l,m_l,j$ are the orbital, projection and total angular momentum quantum numbers and $Y_{l,m_l}(\theta,\phi)$ are spherical harmonic functions. However, since a plane wave is polarized transversely, we can omit $\widehat{z}$ operators. Therefore, in the Faraday geometry, where the $k$-vector of the light is parallel with the magnetic field along $z$, we have:

\begin{equation}
   \mathcal{M}_{\rm{F}}= \sqrt{\frac{4\pi}{3}}\langle l'm_l'|\frac{\widehat{x} + \mathrm{i}\widehat{y}}{\sqrt{2}}Y_{1,1}+\frac{\widehat{x} - \mathrm{i}\widehat{y}}{\sqrt{2}}Y_{1,-1}|lm_l\rangle.
\end{equation}

\noindent Given that $\Delta m_l = \Delta m_J$ since $\Delta m_s = 0$, left/right handed circular light induces $\sigma^{+/-}$ transitions in the Faraday geometry \cite{adams2018}.
\\\\\
In the Voigt geometry, the $k$-vector of the light is perpendicular to the magnetic field. Without loss of generality, we choose to rotate in the $x-z$ plane. As such, we substitute $\widehat{z}\rightarrow \widehat{x}$, $\widehat{x}\rightarrow -\widehat{z}$ in eq. \ref{eq}. This gives the Voigt angular dipole matrix element:

\begin{equation}
   \mathcal{M}_{\rm{V}}=\sqrt{\frac{4\pi}{3}}\langle l'm_l'|\widehat{x}~Y_{1,0}+\frac{\mathrm{i}\widehat{y}}{\sqrt{2}}(Y_{1,1}-Y_{1,-1})|lm_l\rangle.
\end{equation}

\noindent Therefore, horizontally polarized light induces $\pi$ transitions and vertically polarized light induces a linear combination of $\sigma^+$ and $\sigma^-$ transitions \cite{Voigt4}. The reader can derive the expression if we instead rotate in the $y-z$ plane.
\\\\\
A general expression for all geometries is achieved by rotating the magnetic field by an angle $\theta$. This gives the quantization axis transformations as:
\begin{equation}
\begin{split}
 \widehat{x} \rightarrow \widehat{x}\cos\theta-\widehat{z}\sin\theta, \\
 \widehat{z} \rightarrow \widehat{z}\cos\theta+\widehat{x}\sin\theta.
\end{split}
\label{eq4}
\end{equation}

\noindent We then arrive at the general angular dipole matrix element by applying the transformations in Eq. \ref{eq4} to Eq. \ref{eq}:

\begin{equation}
\begin{split}
   \mathcal{M}_{\rm{O}}= \sqrt{\frac{4\pi}{3}}\langle l'm_l'|&\frac{\widehat{x}\cos\theta+ \mathrm{i}\widehat{y}}{\sqrt{2}}Y_{1, 1}+\frac{\widehat{x}\cos\theta- \mathrm{i}\widehat{y}}{\sqrt{2}}Y_{1,- 1}\\&+\widehat{x}\sin\theta Y_{1,0}|lm_l\rangle.
   \end{split}
\end{equation}

\noindent If the magnetic field is oblique to the $k$-vector, the projection operators are elliptical for $\sigma$ transitions and horizontally linear for $\pi$ transitions. However, the operators are non-orthogonal. Therefore, at least two transitions will be induced for any given input polarization (provided the probability of the transition in question is non-zero at the particular frequency).

\section*{Appendix B: Calculating Invariant Polarization States in the Oblique Geometry }
Following the Jones Calculus method from \cite{rotondaro}, the propagation eigenmodes $\vec{a}$ and $\vec{b}$ form the conjugate transposed ($^\dagger$) rotation matrix,

\begin{equation} \mathbf{R} = \begin{bmatrix} a_1 & b_1 \\ a_2 & b_2 \end{bmatrix}^\dagger,
\label{eigenmode definition}
\end{equation}

\noindent such that the input electric field $\mathbf{E_{\rm{in}}}$ is propagated to the output electric field $\mathbf{E_{\rm{out}}}$ as

\begin{equation} \mathbf{ E_{\rm{out}} = R^{-1} \cdot T \cdot R \cdot  E_{\rm{in}}},
\label{prop2}
\end{equation}

\noindent where $\mathbf{T}$ is a diagonal matrix with first and second diagonal terms $f(n_a)$ and $f(n_b)$, the function $f$ applied to the refractive indices $n_a$ and $n_b$. These refractive indices are the eigenvalues associated with $\vec{a}$ and $\vec{b}$ respectively.
\\\\
If an invariant polarization $\vec{p}$ or $\vec{q}$ is the input field to eq. \ref{prop2}, the output field will only depend on one of $f(n_a)$  or $f(n_b)$ . As such, no phase difference is induced and the polarization state does not evolve.
\\\\
To calculate the invariant polarizations, we can choose a state orthogonal to an eigenmode. Formally, we say that the invariant polarizations and eigenmodes together form a \textit{biorthogonal} system \cite{brody2013biorthogonal}.
\\\\
The need for a biorthogonal system is a direct result of a non-Hermitian system Hamiltonian \cite{berry2004physics,bender2007making,berry2011optical,el2019dawn,el2018non}. The eigenvalues (refractive indices) $n_a$ and $n_b$ are in general complex.
\\\\
In the Faraday and Voigt cases, the eigenmodes are orthogonal. Therefore, the invariant polarizations are the eigenmodes and we say the system is \textit{orthonormal} provided the eigenmodes are normalized. 
\\\\
We prove this in the Faraday case where the eigenmodes are left hand and right hand circular light. We input an eigenmode, left hand circular light,

\begin{equation} 
\begin{split}
\mathbf{ E_{\rm{out}}} &= \frac{1}{2} \begin{bmatrix}  1 & 1 \\ -\mathrm{i} & \mathrm{i}  \end{bmatrix}\cdot \begin{bmatrix} {{f(n_a)}} & 0 \\ 0 & {f(n_b)} \end{bmatrix}\cdot \overbrace{\begin{bmatrix} 1 & -\mathrm{i} \\ 1 & \mathrm{i} \end{bmatrix}}^{\mathlarger{\mathlarger{\mathlarger{\mathlarger{ \circlearrowleft, \circlearrowright}}}}} \cdot \overbrace{\begin{bmatrix} 1 \\ \mathrm{i} \end{bmatrix}}^{\mathlarger{\mathlarger{\mathlarger{\mathlarger{\circlearrowleft}}}}} \\
&= \frac{1}{2} \begin{bmatrix} {{f(n_a)}}+{f(n_b)} & -\mathrm{i}({f(n_a)}-{f(n_b)}) \\ \mathrm{i}({f(n_a)}-{f(n_b)}) & {{f(n_a)}}+{f(n_b)}  \end{bmatrix} \cdot \begin{bmatrix} 1 \\ \mathrm{i} \end{bmatrix}
\\
&= \begin{bmatrix} f(n_a) \\ \mathrm{i} \cdot f(n_a) \end{bmatrix} \\
&= f(n_a) \cdot \begin{bmatrix} 1 \\ \mathrm{i} \end{bmatrix} \begin{matrix} \vphantom{x}\\ \vphantom{y}.\end{matrix}
\end{split}
\label{faraday_example}
\end{equation}

\noindent The final output field is the same as the input field multiplied by a factor so the eigenmode is also an invariant. Similarly, we prove this in the Voigt case where the eigenmodes are horizontally and vertically linear light. We input horizontally linear light,

\begin{equation} 
\begin{split}
\mathbf{ E_{\rm{out}}} &= \begin{bmatrix}  1 & 0 \\ 0 & 1  \end{bmatrix}\cdot \begin{bmatrix} {{f(n_a)}} & 0 \\ 0 & {f(n_b)} \end{bmatrix}\cdot \overbrace{\begin{bmatrix} 1 & 0 \\ 0 & 1 \end{bmatrix}}^{\longleftrightarrow, \hspace{0.1cm}  \updownarrow} \cdot\overbrace{\begin{bmatrix} 1\\ 0 \end{bmatrix}}^{\longleftrightarrow} \\ 
&= \begin{bmatrix} f(n_a) & 0) \\ 0 & f(n_b) \end{bmatrix} \cdot \begin{bmatrix} 1 \\ 0 \end{bmatrix} \\
&= \begin{bmatrix} f(n_a) \\ 0 \end{bmatrix} \\
&= f(n_a) \cdot \begin{bmatrix} 1 \\ 0 \end{bmatrix}
\end{split}
\end{equation}

\noindent Like the Faraday case, the output field is the input field multiplied by a factor and so the eigenmode is also an invariant polarization.
\\\\
In the oblique case, the eigenmodes and the invariant polarizations are \textit{not} one and the same since the oblique eigenmodes are in general non-orthogonal. Without loss of generality, we choose the left hand orthogonal vectors to the eigenmodes as our oblique invariant polarizations,

\begin{equation}
\vec{p} = \begin{pmatrix} - a_2\\ a_1 \end{pmatrix}^*, \hspace{0.4cm}
 \vec{q} = \begin{pmatrix} - b_2 \\ b_1 \end{pmatrix}^*,
 \label{invariant_transf}
\end{equation}
\noindent with $^*$ indicating the complex conjugate.
\\\
\setlength{\belowcaptionskip}{0.5cm}

\begin{figure}[!tb]
\centering
{\includegraphics[width=0.8\linewidth]{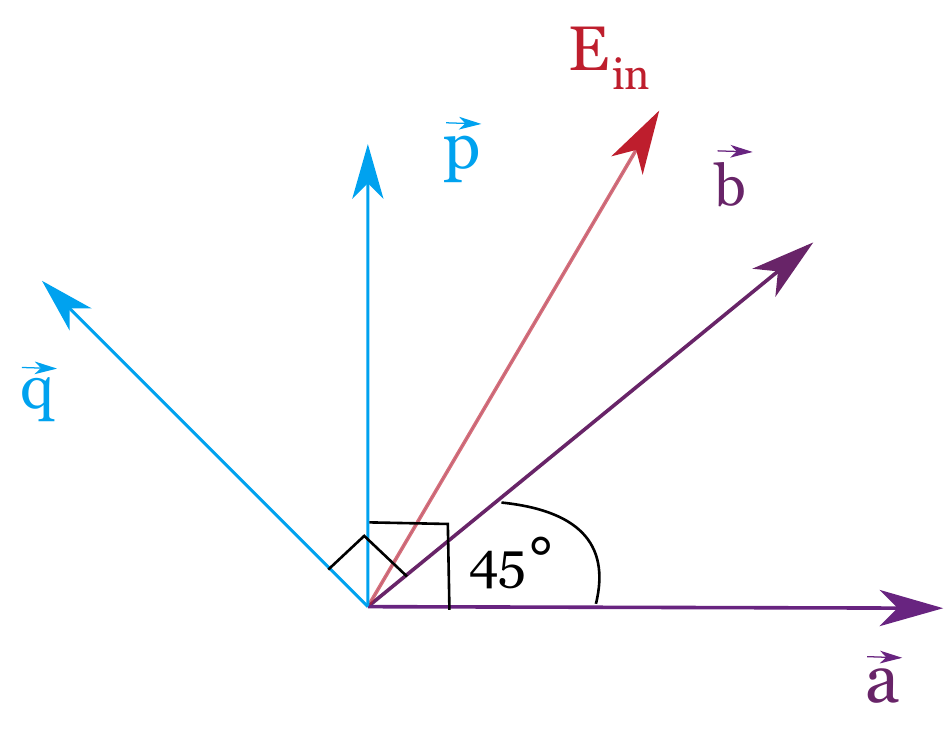}}
\caption{A diagram of the vectors used in eq. \ref{matrix eq34}. In purple, we have the linear horizontal eigenmode, $\vec{a}$, and the linear diagonal eigenmode at $45^\circ$, $\vec{b}$. In blue, the invariant polarizations to $\vec{a}$ and $\vec{b}$, $\vec{p}$ and $\vec{q}$ respectively. In red, an arbitrary input polarization, $\mathbf{E_{\rm{in}}}$, to be propagated through the vapor.}
\label{toy-example figure}
\end{figure}
\noindent We illustrate this situation using a toy expression. Fig. \ref{toy-example figure} is a diagram of the relevant eigenmodes and invariant polarizations. Our eigenmodes will be two non-orthogonal polarizations: horizontally polarized light and linear light at 45$^\circ$. We input linear horizontally polarized light. Eq. \ref{prop2} then gives,

\begin{equation}
\begin{split}\mathbf{ E_{\rm{out}}}&=\begin{bmatrix} 1 & 0 \\ -1 & 1\\ \end{bmatrix}\cdot \begin{bmatrix} {{f(n_a)}} & 0 \\ 0 & {f(n_b)} \end{bmatrix}\cdot \overbrace{\begin{bmatrix} 1 & 0 \\ 1 & 1\end{bmatrix}}^{\longleftrightarrow, \hspace{0.1cm} \neswarrow} \cdot \overbrace{\begin{bmatrix} 1 \\ 0  \end{bmatrix}}^{\longleftrightarrow} \\ &= \begin{bmatrix} {{f(n_a)}} & 0\\ {f(n_b)}-{f(n_a)} & {f(n_b)} \end{bmatrix} \cdot \begin{bmatrix} 1 \\ 0 \end{bmatrix} \\
&=\begin{bmatrix} {f(n_a)}\\ {f(n_b)}-{f(n_a)} \end{bmatrix}  \\
&= {f(n_b)} \cdot \begin{bmatrix} 0\\ 1  \end{bmatrix}-{f(n_a)} \cdot \begin{bmatrix} -1 \\ 1 \end{bmatrix}\\
 &= {f(n_b)} \cdot \updownarrow -{f(n_a)} \cdot \nwsearrow.
\end{split}
\label{matrix eq34}
\end{equation}

\noindent While we have input an eigenmode, since the eigenmodes are not orthogonal, the polarization is not invariant. We can decompose the result into two invariant polarizations, linear diagonal light at $-45^\circ$ and vertically polarized light. Note the relationship between the eigenmodes and the invariant polarizations is as described in eq. \ref{invariant_transf}.
\\\\
Into the same system, we input an invariant polarization, vertically linear light, and prove its invariance,

\begin{equation}
\begin{split}
\mathbf{E_{\rm{out}}} &= \begin{bmatrix} {{f(n_a)}} & 0\\ {f(n_b)}-{f(n_a)} & {f(n_b)} \end{bmatrix} \cdot \overbrace{\begin{bmatrix} 0 \\ 1 \end{bmatrix}}^{\updownarrow}\\
&=\begin{bmatrix} {0}\\ {f(n_b)} \end{bmatrix}  \\
&= {f(n_b)} \cdot \begin{bmatrix} 0\\ 1 \end{bmatrix}
\end{split}
\end{equation}

\noindent Of importance to emphasize is this is a toy example. Horizontally linear light and linear light at $45^\circ$ could never jointly form the eigenmodes of a system. This is because $\mathbf{R}$ is  a rotation matrix for some angle $\beta$ and multiplicative factor $\mathcal{F}$ implying a constraint,

\begin{equation} 
\begin{split} \mathbf{R} =\begin{bmatrix} a_1 & b_1 \\ a_2 & b_2 \end{bmatrix}^\dagger = \mathcal{F} \cdot \begin{bmatrix} \cos\beta & -\sin\beta \\ \sin\beta & \cos\beta \end{bmatrix} \\
\implies \lvert a_1\cdot b_1 \rvert = \lvert a_2\cdot b_2 \rvert.
\label{constraint}
\end{split} \end{equation}

\noindent The eigenmodes of the Faraday and Voigt geometry respectively adhere to this constraint. The constraint implies elliptical eigenmodes in the oblique geometry.

\section*{Appendix C: Equivalent Propagation of both passes in the Double Pass Filter}

We consider first the different angles between magnetic field and propagation direction of both passes, $\alpha$ and $\alpha^* = (180- \alpha)^\circ$. Equivalently, we can say in the second pass the magnetic field makes an angle $\alpha$ with a vector pointing in the opposite direction to $\vec{k}$. 
\\\\
In this picture, the relevant axis for identifying polarizations has flipped. Instead of looking at the electric field towards the source, we should view it looking from the source. Therefore, left/right handed polarizations now appear to be right/left handed.
\\\\
Given that everything else remains identical, we can say that the evolution of a left/right handed wave with magnetic field at angle $\alpha$ to the $k$-vector is the same as the evolution of a right/left handed wave with magnetic field at angle $\alpha^*$ to the $k$-vector. If linearly polarized light is input, the filter transmission will not differ in the two cases.
\\\\
The filter output from a horizontally ($\mathbf{E_{\leftrightarrow}}$) or vertically ($\mathbf{E_{\updownarrow}}$) polarized input with crossed polarizers is always the same regardless of geometry,

\begin{equation} ||\mathbf{P_{\updownarrow} \cdot R^{-1} \cdot T \cdot R \cdot  E_{\leftrightarrow}}|| = ||\mathbf{P_{\leftrightarrow} \cdot R^{-1} \cdot T \cdot R \cdot  E_{\updownarrow}}||,
\label{equiv}
\end{equation}

\noindent where $\mathbf{P_{\updownarrow}}$ and $\mathbf{P_{\leftrightarrow}}$ are vertically and horizontally linear polarizers respectively. This follows directly from eq. \ref{constraint}. The two facts stated above explain why both passes of the double pass filter have the same filter transmission.


\end{document}